\documentclass{raa_twocolumn}  

\usepackage{graphicx,times}             %for PS/EPS graphics inclusion, new
\usepackage{natbib}
\usepackage{amssymb,amsmath}
\bibpunct{(}{)}{;}{a}{}{,}
\usepackage[pagebackref=true]{hyperref}

\begin{document}

  \title{ECLAIRs: the SVOM high-energy transient trigger camera}
%   \subtitle{I. Place Your Subtitle Here}

   \volnopage{Vol.0 (2026) No.0, 000--000}      %%preserved for Editor. DOn't remove!
   \setcounter{page}{1}          %%starting page, preserved for Editor. DOn't remove!

   \author{O. Godet
      \inst{1,*}\footnotetext{$*$Corresponding Authors, these authors contributed equally to this work.}
   \and J.-L. Atteia 
      \inst{1}
   \and S. Schanne 
      \inst{3}
    \and C. Lachaud 
      \inst{2}  
   \and A. Goldwurm 
      \inst{2}
   \and F. Piron 
      \inst{4}
    \and Ph. Guillemot 
    \inst{7}    
    \and C. Amoros
     \inst{1}
    \and W. Bertoli
      \inst{2}  
    \and L. Bouchet 
      \inst{1}
    \and M. C. Charmeau
     \inst{7}  
    \and F. Château
      \inst{3}  
    \and B. Cordier 
      \inst{3}
    \and N. Dagoneau 
      \inst{3}
    \and F. Daly
      \inst{3}
     \and J.-P. Dezalay
      \inst{1}
    \and J. Galezzi 
     \inst{7} 
    \and A. Givaudan
      \inst{2} 
    \and A. Gros
      \inst{3}  
    \and M. Karakac
      \inst{2} 
    \and K. Lacombe
      \inst{1}   
    \and H. Leprovost 
      \inst{3}
    \and S. Maestre
      \inst{1} 
    \and K. Mercier
      \inst{7} 
    \and H. Pasquier
      \inst{7} 
    \and L. Perraud
      \inst{7}   
    \and R. Pons
      \inst{1}
    \and D. Rambaud
      \inst{1} 
    \and O. Simonella
      \inst{7}  
    \and T. Tourrette 
      \inst{3}
    \and H. Triou
      \inst{3} 
    \and V. Waegebaert
     \inst{1}
    \and P. Bacon 
     \inst{2}
    \and T. Barlyaeva 
    \inst{4}
    \and N. Bellemont 
     \inst{2}
     \and M.-G. Bernardini 
    \inst{4,6}
    \and M. Brunet
    \inst{1}  
   \and F. Cangemi 
      \inst{2}
       \and C. Cavet 
     \inst{2} 
     \and A. Coleiro
    \inst{2}  
    \and D. Corre 
     \inst{5} 
    \and F. Daigne 
      \inst{5}
    \and A. Foisseau
     \inst{2}
    \and O. Gevin
    \inst{3}
    \and S. Guillot 
      \inst{1}
    \and U. Jacob
    \inst{4} 
    \and F. Lacreu
    \inst{5}
    \and S. Le Stum 
      \inst{2}
    \and P. Maeght 
    \inst{4}
    \and T. Maiolino
    \inst{4}
    \and A. Maïolo 
    \inst{4}
    \and G. Tcherniatinksy 
    \inst{5}
    \and J. Wang
    \inst{5}
    \and H. Yang
    \inst{1}
}

   \institute{IRAP, Université de Toulouse/CNRS/CNES, 9 avenue du colonel Roche, 31028 Toulouse, France; {\it ogodet@irap.omp.eu}\\
%% Please give the E-mail address of the author, to whom future correspondence
%% requests will be sent.
        \and
             Université Paris Cité, CNRS, CEA, Astroparticule et Cosmologie, F-75013 Paris, France;\\
        \and
            CEA Paris-Saclay, Institut de Recherche sur les lois Fondamentales de l'Univers, 91191 Gif sur Yvette, France; \\
        \and  
             Laboratoire Universe et Particules de Montpellier, CNRS/IN2P3, Place Eugène Bataillon—CC 72, Montpellier, France;\\
        \and 
             Sorbonne Université, CNRS, UMR 7095, Institut d’Astrophysique de Paris, 98 Bis bd Arago, 75014 Paris, France;\\ 
        \and   
        INAF–Osservatorio Astronomico di Brera, via E. Bianchi 46, I–23807 Merate, Italy;\\
        \and 
        CNES, 18 Avenue Edouard Belin, 31401 Toulouse cedex 9, France\\
\vs\no
   {\small Accepted 2026 March}}

\abstract{
The core instrument of the {\it SVOM} Gamma-ray burst mission launched in June 2024 is the 4--150~keV 2-D coded mask camera ECLAIRs responsible for the autonomous trigger and localization of transient events
within its field of view. The flight model of ECLAIRs has been built by several French labs (IRAP, CEA, APC) under the supervision of the French Space Agency (CNES), while APC, LUPM and IAP built a suite of data reduction and analysis softwares. This paper outlines the main science goals of ECLAIRs and describes the different instrument sub-systems and their main characteristics. The paper then discusses the instrument configuration and operation as well as the main in-flight measured performances. Finally the paper summarizes the science performance of ECLAIRs up to March 31, 2025.    
}

   \authorrunning{O. Godet et al.}            %author_head in even pages
   \titlerunning{ECLAIRs HE transient trigger camera}  % title_head in odd pages

   \maketitle
%% The author head (on even pages) and the title head (on odd pages) will be
%% automatically extracted from \author{} and \title{}. Whenever the title is too long,
%% you will be asked to supply a shorter one by inserting either \authorrunning{} or
%% \titlerunning{} before \maketitle. Anyway, you can specify your own heads.
%%
%%
%% Note: In the following text body of your manuscript, please note several differences from
%%       other major journals:
%% (1) \subsection{Please Capitalize the First Letter of Each Notional Word in Subsection Title}
%% (2) Please Capitalize the First Letter of Each Notional Word in all tables' captions
%
%________________________________________________ sections below
%
 
\section{Introduction}           %% first-level sections will be auto-capitalized
\label{sect:intro}

ECLAIRs (heareafter ECL -- \citealt{Godet14}) is the prime instrument onboard the Chinese-French Space-based multi-band Variable Object Monitor ({\it SVOM}) mission launched in June 22 2024 \citep{Wei16, Cordier25}. {\it SVOM} is dedicated to the study of gamma-ray bursts (GRBs -- e.g. \citealt{Zhang07}) and other high-energy transients. ECL is a 4--150 keV coded-mask camera in charge of autonomously detecting new transient events within its wide field of view (FoV $\sim 2$ sr) and providing their first localization with an accuracy better than 13 arcmin (90\% confidence level -- c.l.) in near real time.
Once ECL detects a new transient above a certain threshold (slew threshold), the spacecraft promptly slews -- when possible -- to place the source within the FoV of the 0.3--10 keV Microchannel X-ray Telescope (MXT; FoV$_\mathrm{MXT}$ = $58\times 58$ arcmin$^2$ -- \citealt{MXT25}) and the Visible Telescope (VT; FoV$_\mathrm{VT}$ = $26\times 26$ arcmin$^2$ -- \citealt{VT25}) in order to further refine the source position and to monitor its flux evolution. The spectral characterization of the GRB prompt emission by ECL is complemented by the Gamma-Ray Monitor (GRM -- \citealt{GRM25}) extending the energy coverage up to 5 MeV.

ECL has been built by three French institutes (APC, CEA \& IRAP) with a science lead by IRAP and under the supervision of the French Space Agency (CNES). The data reduction and analysis pipelines have been developed by APC \citep{Goldwurm25} as well as LUPM and IAP \citep{Piron25}.   

In Section~\ref{sect:goals}, we recall the main science drivers for the ECL instrument. The description of the instrument is outlined in Section~\ref{sect:data}. For more details on the ECL sub-systems, refer to \cite{Godet25} for the detection plane and the readout electronics, \cite{Lachaud25} for the coded mask and \cite{Schanne25} for the onboard software including the trigger algorithms implemented in the data processing unit. In-flight calibration results gathered mainly during the commissioning phase are discussed in Section~\ref{sect:perfo} showing that the instrument performs overall very well and displays excellent performance. In Section~\ref{sect:science}, we finally present some science highlights for both the GRB core program and observatory science, outlying how ECL helps unveiling new populations of high energy (HE) transients thanks to its unique characteristics and performance.

\section{Science goals}
\label{sect:goals}

Since its conception, \textit{SVOM} has been designed to detect all types of GRBs and characterize them in details. 
This choice has driven many important characteristics of the mission described in this volume, like the pointing strategy, the fast-slewing capability, the selection of on-board instruments, a dedicated scientific ground segment, etc.

At the instrument level, this choice had various impacts on the design of ECL: i) the most significant being the extension to low energies (down to 4\,keV) of the instrument energy range; ii) the implementation of a highly flexible source trigger algorithm; iii) the decision to send all the recorded photons to the ground.
This strategy has proven fruitful with the detection of 32 GRBs (29 onboard and 3 on-ground) by ECL up to March 31, 2025, among them a number of classical long (or very long lasting for a few minutes) GRBs, two short-hard GRBs, a sample of bona-fide X-ray flashes, a soft GRB possibly due to a supernova (SN) shock breakout. 
These cosmic transients span a broad range of redshifts, from z = 0.238 to 7.3 \cite{Cordier25}, and isotropic equivalent energies, from $10^{50}$  to $5 \times 10^{53}$ erg. 

This comprehensive sample contains promises of significant progress in our understanding of the GRB phenomenon and their use for cosmological studies, thanks to the observation of ECL-detected GRBs with the unique suite of {\it SVOM} instruments, combined with their efficient follow-up by satellites, including the Neil Gehrels {\it Swift} observatory, {\it Einstein Probe} in X-rays and {\it JWST} in infra-red (IR), with several robotic observatories (mostly in optical), and with large telescopes equipped with spectrographs (e.g. Very Large Telescopes, Nordic Optical Telescope, Gran Telescopio Canarias).
Specific ways of progress thanks to ECL include: i) The detection of some very long GRBs, lasting sufficiently ($>$ 50--100\,s) to permit the observation of their prompt phase with narrow-field instruments after satellite slews; ii) A confrontation of the diversity of the prompt emission with the properties of the optical and X-ray afterglows and the possible hosts, to get a detailed characterization of these various types of explosions and their possible origin; iii) Detailed studies of the poorly known population of soft faint transients, in synergy with {\it Einstein Probe}; iv) Finding electromagnetic counterparts to gravitational wave and neutrino events as well as GRB associations to supernova/kilonova, to get deeper insights in the progenitor nature and properties -- see also \cite{Daigne25}.

Within the \textit{SVOM} mission, ECL not only detects and localizes cosmic transients, but it also monitors the hard X-ray sky -- see \cite{Coleiro25} and Section~\ref{sect:science}. 
The transmission of all the photons detected by ECL to the ground allows monitoring the activity of known Galactic and extra-galactic X-ray sources in the 4 -- 150 keV energy range including the detection of a burst from the Soft Gamma-ray repeater SGR 1806--20, searching for transients undetected on-board, and performing targeted searches for sub-threshold events with known time and position (e.g. neutrino and gravitational wave alerts, transients detected at other wavelengths such as fast radio bursts, supernovae, fast blue optical transients, tidal disruption events, blazars, etc.) -- see \cite{Brunet26}. This also allows studying some Solar system or Earth phenomena (e.g. terrestrial gamma-ray flashes as well as solar activity events such as Solar flares and possibly X-ray auroras).

\section{Instrument description and configuration}
\label{sect:data}

Onboard {\it SVOM}, ECL is in charge of autonomously detecting in real time GRBs and other high energy transients within its FoV and providing their first localization as well as characterizing the GRB prompt emission both temporally and spectrally. Thus, ECL is a wide FoV ($\sim 2$ sr) coded mask spectro-imaging instrument operating in photon counting mode in the 4 -- 150 keV band for a total mass of 89.6 kg. Setting the low energy threshold down to 4 keV to enhance the ECL sensitivity to soft gamma-ray bursts and high-redshift GRBs (\citealt{Godet09}) has major impacts on the instrument design as described below.

 \begin{figure*}
        \centering
        \includegraphics[width=1\linewidth]{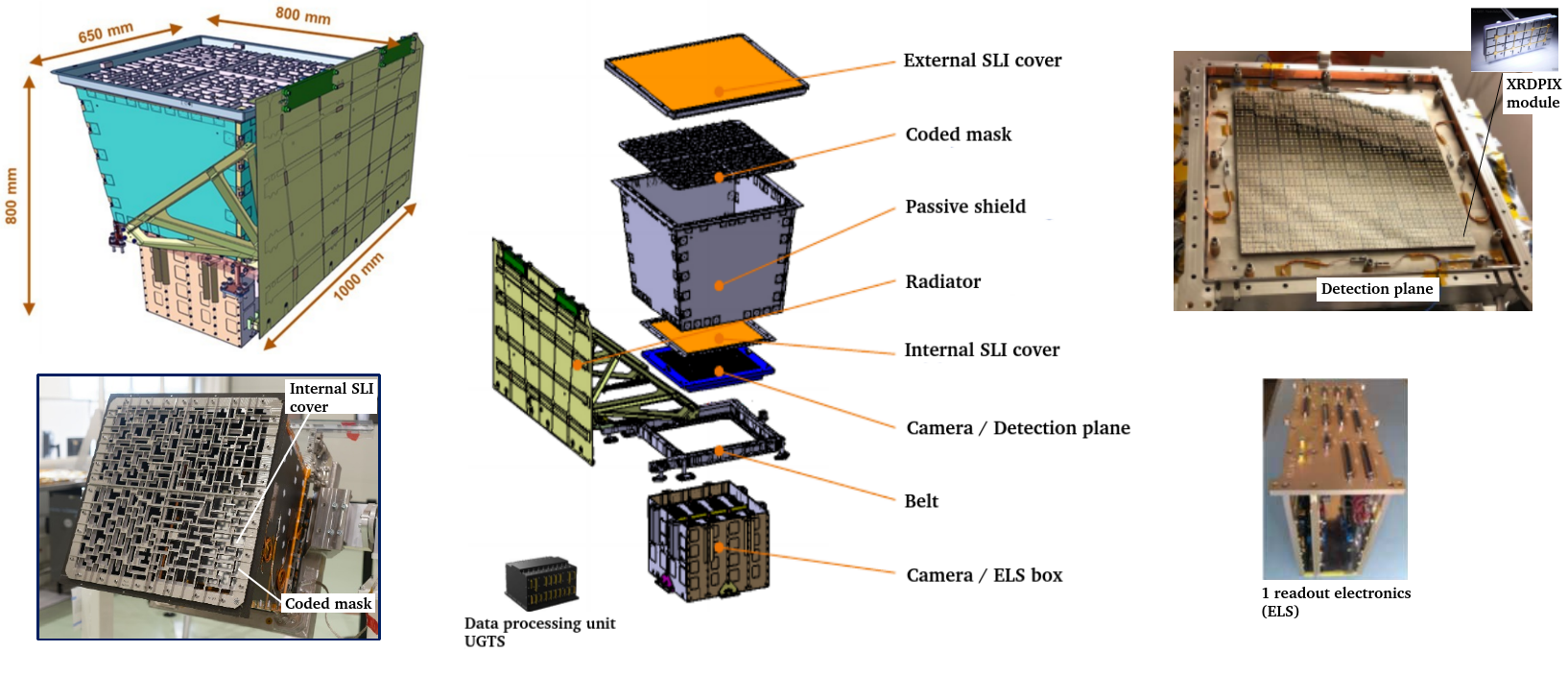}
        \caption{The ECL instrument and its main subsystems. SLI covers helps with the shield to create a cavity that is opaque to optical light to avoid optical loading on the detectors. The detection plane is paved with 200 detection modules called XRDPIX, each of them made of $4\times 8$ Schottky-type CdTe pixels.}
        \label{fig:eclairs}
    \end{figure*}

The 2-D coded mask imaging technique (\citealt{Goldwurm22}) relies in projecting the source light passing through a 2-D coded mask made of opaque and transparent elements onto a pixelated detection plane built as an array of $80 \times 80$ Schottky-type CdTe detectors placed at around 46 cm below. The mask pattern is chosen so that the projected mask shadow onto the detection plane (shadowgram) is unique for any source position within the ECL FoV. Deconvolving the recorded shadowgram by the mask pattern allows to retrieve the source positions within the FoV. The contribution from bright sources within the ECL FoV can also be subtracted before performing the deconvolution to avoid coding noise induced by these bright sources in the sky images (see \citealt{Schanne25}).         
Table~\ref{Tab1} summarizes the main characteristics of ECL and Figure~\ref{fig:eclairs} shows the main parts of the flight model. Below we provide a description of the main sub-systems of ECL.  

\begin{table}
\begin{center}
\caption[]{ECLAIRs main science characteristics and performances.}\label{Tab1}
 \begin{tabular}{ll}
  \hline\noalign{\smallskip}
Parameter &  Characteristic                 \\
  \hline\noalign{\smallskip}
Detectors  &  6400 Schottky-type CdTe \\
           &  detectors ($4\,\times\,4\,\mathrm{mm}^2$\\
           &  and 1 mm thick) \\
Mask aperture fraction & 40\% \\
Field of view (total) & 2.05 sr ($89^\circ \times 89^\circ$) \\ 
Fully coded FoV & ($23^\circ \times 23^\circ$) \\
Time resolution & 20 $\mu$s \\
Data acquisition & Photon single/multiple\\
                 & interaction tagging \\
Energy range   & 4 -- 150 keV  \\ 
Data rate & $\le 18$ Gb/day \\
 \hline
 \hline
Geometrical area  &    $\sim 950~\mathrm{cm}^2$                 \\
of the detection plane & \\
Total effective area [10--70 keV]  & $\ge 340~\mathrm{cm}^2$        \\
Photopeak effective area @ 6 keV & $\ge 200~\mathrm{cm}^2$ \\
Sensitivity to a long GRB over  & $2.5\times 10^{-8}$\,erg\,cm$^{-2}$\,s$^{-1}$\\
1\,s in 5 -- 50 keV & \\

Source positioning error   & $< 11.5$ arcmin  \\
    (90\% c.l.*)          & for sources with SNR\,$>\,8$ \\

Angular resolution (on-axis) & 1.56 deg \\

Dead time & $< 10\,\%$ for $10^5$ c/s  \\

Energy calibration accuracy & $\le 0.3$ keV below 80 keV \\
Energy resolution @ 60 keV & $<1.2$ keV \\

  \noalign{\smallskip}\hline
  * confidence level
\end{tabular}
\end{center}
\end{table}

\subsection{Detection plane}

Built and calibrated by IRAP, the detection plane (\citealt{Godet25}) is made of 200 elementary detection modules so-called XRDPIX (\citealt{L13}) mounted onto an AlBeMet cold plate, totaling a geometrical surface of about 950 cm$^2$. Each XRDPIX module is made of a $8\times 4$ matrix of $4\times 4$ mm$^2$ and 1\,mm-thick Schottky-type CdTe detectors from Acrorad Co., Ltd. (Japan) hybridized with the CEA low-noise and low power consumption ASIC IDeF-X (\citealt{Gevin09}). The detectors are cooled down to --20°C and reverse-biased at --300 V (see \citealt{Godet25}). For redundancy purpose, the detection plane is divided into 8 sectors, each gathering 25 XRDPIX modules. Each sector is connected to an almost electrically independent readout electronics (Electronique Secteur -- ELS). 
Events detected above the low-energy threshold are time-tagged at a 20 $\mu$s time resolution, position-tagged onto the plane and their energy is encoded over a 10 bit dynamics if the event is classified as a single or double event (see below). Saturating events are encoded in channel 1023. 
The readout electronics performs event classification flagging the events as single or multiple events to discriminate photons depositing all their energy in one detector (single events) from other events induced by fluorescence/Compton interaction processes or particle showers hitting several pixels within some time coincidence windows. We chose to encode the energy for double events since they carry information about fluorescence and Compton interactions within detectors that is used for in-flight calibration activities (Cd and Te fluorescence lines) and may be useful for performing some polarization measurements of bright sources observed by ECL. The ELS also count the global event counts per pixel before event classification over a 8 s time frame (the Time Frame Events -- TFE). 

The ELS tag the single event frames into four adjustable energy bands in keV (EBANDS = [5 -- 8], [8 -- 20], [20 -- 50], [50 -- 120]) used by the trigger algorithms to build shadowgrams in different energy strips in keV (ESTRIP = [8 -- 120], [8 -- 50], [5 -- 20], [5 -- 8]) (see below).  
The readout electronics is designed to support an event flow from an extremely bright source up to  $5 \times 10^4$ events s$^{-1}$ over the whole detection plane with a dead time of less than 5\%. 
For more detail on the detection plane, see \cite{Godet25}.

\subsection{Coded mask}

Built by APC, it has an aperture fraction of 40\% and it makes use of a quasi-random pattern. Because of the 4 keV low-energy threshold, the mask could not be supported by any material. Instead to mitigate thermo-mechanical deformation, the mask has been pre-constrained and tightened with a cross. The mask is $54\times 54$ cm$^2$ and it is made of a 0.6 mm thick tantalum sheet sandwiched by two titanium sheets in order to ensure its flatness and hardness. The thickness of the tantalum sheet offers a $> 99\%$ opacity of the mask elements below 50 keV. The titanium sheets have been machined in such a way as to reduce the mask weight and the vignetting impacts for off-axis sources. For more detail on the coded mask, see \cite{Lachaud25}.

\subsection{Shield \& SLI covers}

{\it Graded shield} -- Built by CNES, it is made of lead (0.8 mm) and copper (0.1 mm) layers supported by an aluminum honeycomb structure covered by a 1.2 mm thick carbon layer. This shield encloses the detection plane and supports the coded mask placed at around 46 cm above the plane; which defines the total FoV of the camera (see Table~\ref{Tab1}). The totally coded FoV is $23\times 23~\mathrm{deg}^2$ (i.e. $\sim 0.15$ sr). The shield ensures a $> 99\%$ opacity for photons coming from outside the instrument FoV with energies less than 50 keV. We make use of the lead (70 -- 90 keV) and copper (8 keV) fluorescence lines for in-flight calibration of the detector energy scale (see Fig.\,6b in \citealt{Godet25} and Section~\ref{sect:nrj}). 

{\it SLI\,\footnote{SLI is a type of multi-layer insulation.} covers} --  In order to prevent optical light to reach the CdTe detectors, in addition to the lateral shield, three SLI covers also built by CNES have been designed to hermetically seal the camera cavity: i) A 25 $\mu$m-thick Kapton layer with a $\sim 1300$ \AA-thick SiOx deposit on one face and a 1000 \AA-thick aluminum deposit on the other. SiOx is there to mitigate SLI degradation by atomic oxygen at the altitude of SVOM ($\sim 625$ km); ii) Two 12.5 $\mu$m-thick Mylar layers with a 1000 \AA-thick aluminum deposit on each face. The latter layers are placed inside the camera cavity and the former one on top of the coded mask with the SiOx deposit facing space (see Fig.~\ref{fig:eclairs}). These covers offer a transparency of $\sim 78\%$ at 4 keV.

\subsection{Data processing unit} 
\label{sect:ugts}
The data processing unit is called UGTS (Unité de Gestion et de Traitement Scientifique). Its main roles are to: i) Create all the voltage powering on the different instrument sub-systems; ii) Manage the instrument configuration and modes, iii) Monitor the instrument health and apply Fault-Detection, Fault-Isolation and Recovery actions in case of onboard issues to be urgently  sorted out; iv) Actively control the detector temperature within its nominal range along the SVOM orbit; v) Manage the acquisition flow of science and housekeeping data; vi) Search for the appearance of new transients within the ECL FoV in real time (trigger) and request the spacecraft slew for follow-up observations; vii) Send all the data produced by ECL to the Payload Data Processing Unit. The UGTS hardware has been designed and built by CNES. The UGTS firmware and onboard software, including
the scientific software (trigger), has been designed, validated and provided by CEA. 

Onboard trigger algorithms -- The search for new transients is performed thanks to two trigger algorithms respectively based on excess searches in systematically built count-rates, confirmed by sky imaging (Count Rate Trigger -- CRT) and excess searches in systematically-built sky images (Image Trigger -- IMT) over different timescales (from 10 ms to 20.48 s for CRT and from 20.48 s to 20 min for IMT) as well as 4 different energy strips. For CRT, the temporal excesses are also searched for into 9 different zones of the detection plane. 

For CRT, once a significant temporal excess has been found above a certain count-rate SNR\,\footnote{Signal-to-noise ratio} threshold in a given energy strip and timescale, a shadowgram is built on this configuration, cleaned for noisy pixels, and a sky image is computed through the deconvolution process. If an excess appears in the sky image above a certain image SNR threshold (typically SNR$^\mathrm{thres}_i = 6.5$) and if the source position is not hidden by the Earth in the FoV, an alert message is issued and promptly sent to the spacecraft as well as to the ground through the VHF network. If the source position corresponds to one of the sources within the onboard catalog (containing the 82 brightest known sources in the ECL energy range -- \citealt{Dagoneau21}), then the alert is of type CAT (catalog source alert), providing the source identification number in the onboard catalog. If this is not the case, the alert is of type GRB candidate. 

For IMT, the sky images in 4 energy strips are built every 20.48 s and stacked together to cover different timescales as mentioned above. Prior to the deconvolution process, the shadowgrams are cleaned for noisy pixels, background inhomogeneities on the detection plane and bright sources present within the FoV at the moment of the observation. Excesses found in the sky images are then treated the same way as for the CRT. 

Since both trigger algorithms run in parallel, CRT and IMT may provide successive alert messages with different detections and localizations over the burst duration. The best source positioning is given by the alert message with the best image SNR from IMT or CRT.
For GRB alerts, if the best source SNR in the sky image is larger than the slew threshold (typically SNR$_\mathrm{slew} = 7$), then the spacecraft will promptly repoint, when possible, MXT and VT towards the best ECL position. This option also exists for CAT alerts but it has not been activated yet.

See \cite{Schanne25} for more detail on the onboard trigger algorithms and alert sequence. 

\medskip
{\bf XOFF/XON mechanism} -- In addition of counting TFE events, the UGTS also records the global number of multiple events per second. This latter global counter provides valuable information on the presence of high energy particles along the {\it SVOM} orbit (\citealt{Claret25}). 
To stay within the instrument daily mass memory allocation of $\sim 18$ Gb per day and to limit the data flow processed by the onboard source trigger algorithms, data collection is inhibited by several mechanisms. First, there is no data collection (including TFE and global multiple count rates) when the satellite crossing time of the deep SAA (South Atlantic Anomaly -- see Section~\ref{sect:SAA}) is longer than 10 minutes. During those times, the detectors are depolarized; which allows to retrieve their full spectroscopic performance when exiting the deep SAA (see \citealt{Godet25}). Secondly, within the extended SAA regions (see Section~\ref{sect:SAA}), if the global multiple count rates are larger than 1600 cts s$^{-1}$, then the XOFF/XON mechanism -- designed to account for unpredicted high background regions along the orbit in real time -- stops data collection (XOFF), except for the global count rates that continue being recorded. During XOFF periods within the extended SAA regions, the trigger algorithms are not operational. When the global multiple count rates fall below 800 cts s$^{-1}$, the data collection resumes (XON) along with the onboard trigger software. Outside the SAA regions, data collection and the onboard trigger software are operational.  
 
Since January 1, 2025, the median fraction of time over which ECL was collecting data is $\sim 86\%$ (which is close to expectations), with a median XON time fraction of $71\%$. This could be accounted by the currently high solar activity inducing high and highly variable particle background noise in the SAA regions (see Section~\ref{sect:SAA}).  This results in a median observing duty cycle of $\sim 61\%$.   

\medskip
{\bf Noisy pixel onboard management} -- Because very noisy pixels could deeply affect the performance of the onboard trigger software, a dedicated algorithm in the UGTS allows to disable at most one very noisy pixel over the whole detection plane per 8\,s time frame when its TFE count rates become larger than 25 cts s$^{-1}$ (i.e. $> 200$ counts over 8\,s -- adjustable number of counts). This limits the cases when several detectors get disabled following a particle shower event hitting the detection plane for instance. Moreover, this algorithm cannot disable more than a certain adjustable number of pixels on the detection plane (set at 640 detectors i.e. 10\% of the plane detectors). The management of noisy pixels is only effective outside the SAA regions. Once disabled, a pixel can no longer be readout by the ELS. In addition, when the satellite exits the deep SAA, automatically disabled pixels are reactivated. If a pixel is systematically disabled when reactivated, we have the possibility on ground to permanently disable it. Prior to the launch, we have identified 10 dead pixels over the 6400 pixels of the detection plane. Since the launch, 4 additional dead pixels were identified; which represents in total $\sim 0.22\%$ of the ECL detectors. In-flight data show that the number of disabled very noisy pixels per day is rather low (i.e. just a few per day). Moreover, we note some abnormal behavior of some modules or columns of modules becoming very noisy for some limited periods of time for some still unknown reasons (\citealt{Godet25}). Similar behavior has been seen during the on-ground calibration campaign. However, ageing of the detectors and/or the electronics does not seem to be the root cause. This results in a massive number of disabled pixels on these noisy modules/columns that led to series of false triggers during the time these pixels stayed disabled. In most cases, once reactivated these pixels perform nominally.

\subsection{Thermal regulation system} 

The thermal regulation of the instrument designed and built by CNES is done through multi-layer insulation covering the whole camera to ensure its global thermal protection and a thermal control system made of CCHPs (C$_2$H$_6$) (Constant Conductance Heat-Pipes) installed on a 0.62 m$^2$ radiator and underneath the cold plate using a NH$_3$ fluid as well as VCHPs (NH$_3$) (Variable Conductance Heat-Pipes) making thermal connection between the cold plate and the radiator through an aluminum interface plate. This is completed by a series of thermal sensors installed on the cold plate and all thermal system elements. The VCHPs are coupled to heaters that are actively controlled by the UGTS through a PID algorithm in order to regulate the detector temperature in their nominal temperature range between --25°C and --18°C with a mean temperature of --20°C. 

\subsection{In-flight configuration}

The parameters values of both the camera (e.g. low energy threshold, EBANDS values, dead pixels, etc.) and the onboard software including the onboard trigger software (e.g. source catalog, mask-to-detector height, XOFF/XON thresholds, etc.) are adjustable through configuration tables. During the commissioning phase, we also adjusted the South Atlantic Anomaly (SAA) contours to improve the instrument operability, to limit the onboard data volume in order to stay within our daily allocation (18 Gb per day) and to optimize its observing duty cycle (see Section~\ref{sect:SAA}). After launch and since then, we adjust, when needed, these various parameters in order to make the instrument more resilient and more efficient. To do so, we use X-band event-by-event data as well as science and engineering housekeeping data. In rare occasions, we updated the onboard software.  All of this is done from the ground as a collaboration from the ECL Instrument Center\,\footnote{The EIC is the centre of expertise of the ECL instrument in charge of monitoring its health and the evolution of its performances, updating the calibration and onboard configuration files as well as the onboard software.} (EIC) for the fine-tuning of the onboard configuration tables as well as for preparing the updates of the onboard software and the CNES French Payload Operation Centre located at Toulouse (France) preparing the telecommands. On-ground validation tests using spare hardware of the UGTS and the detection plane ($1/4$ of the flight model i.e. 1600 detectors) as well as instrument simulators are performed at IRAP, CEA and CNES during the whole process for every updates of the onboard configuration and the onboard software.

\section{In-flight performances}
\label{sect:perfo}

\subsection{Background \& South Atlantic Anomaly}
\label{sect:bkgsaa}

\subsubsection{Background}
\label{sect:bkg}

In the 4 -- 150 keV energy band, the photon background is dominated by the cosmic X-ray background (CXB, \citealt{Churazov07, Ajello08, Turler10}), its reflection by the Earth atmosphere (\citealt{Churazov08}) and atmospheric albedo emission (\citealt{Sazonov07, Ajello08}). 
Although the photon contribution is predictable, the additional background from particles, mainly from the SAA and charged particle precipitations from high to lower altitudes along the {\it SVOM} orbit, is much harder to predict, as their particle flux varies temporally, spatially, and spectrally. The contribution from cosmic rays is negligible. The particle contents within the Earth magnetosphere strongly depends on the solar activity. {\it SVOM} being launched near the solar maximum, the SAA component can dominate the overall background. Particles interact with the atmosphere and the satellite body, generating secondary particles and delayed photon and particle emissions (activation), which makes accurate modeling especially challenging.

\begin{figure}
        \centering
        \includegraphics[width=1.05\linewidth, angle=0, ]{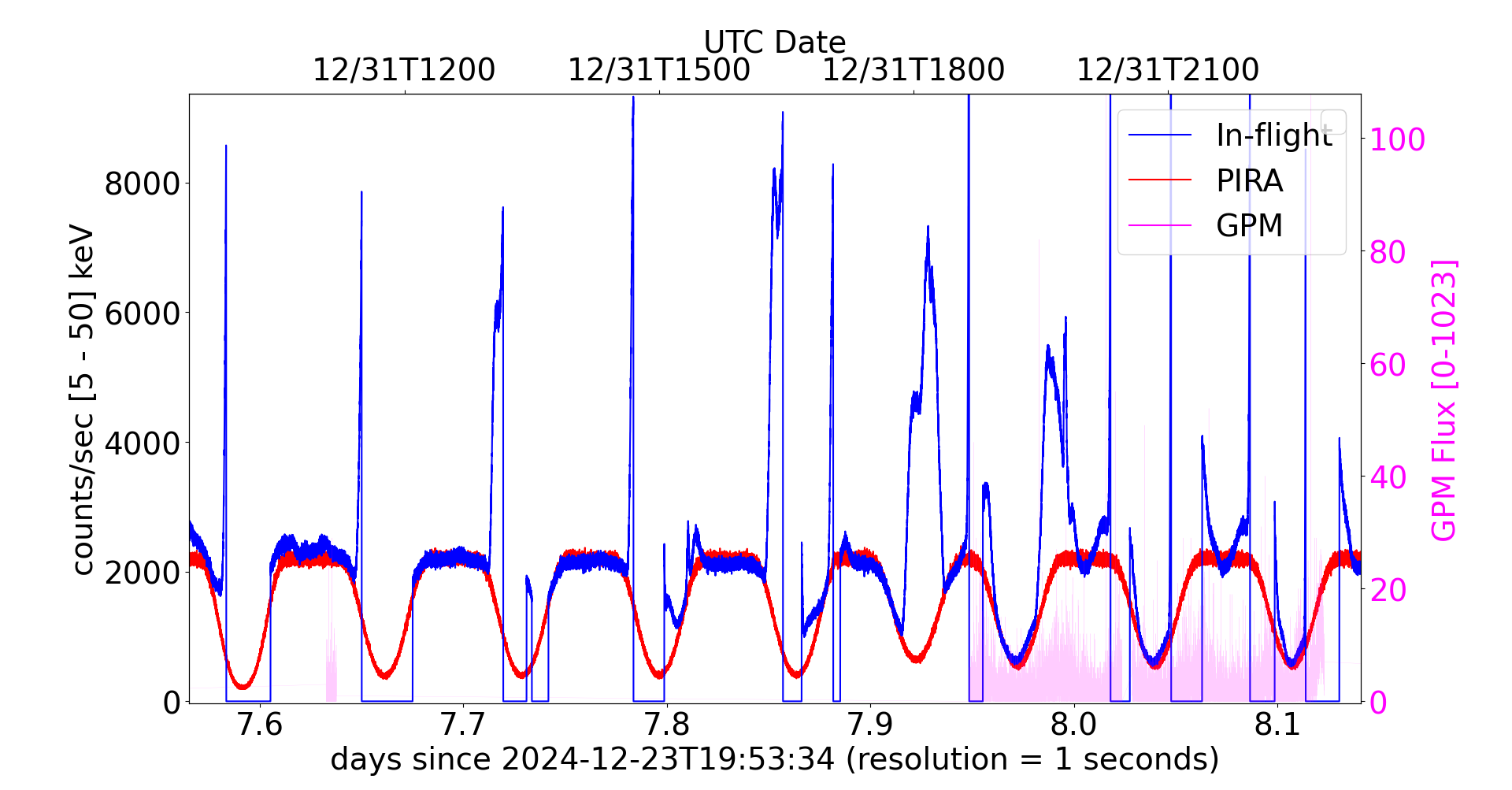}
        \caption{Comparison of the ECL in-flight background lightcurve in the 5 -- 50 keV (blue) with GEANT4-based simulation. The Monte-Carlo predictions include the contributions from the cosmic X-ray background (CXB), the CXB reflection and atmospheric albedo. When available, the count rates from the {\it SVOM}/GRM Particle Monitor (GPM) are shown in pink. High solar activity strongly affects ECL count rates through trapped magnetospheric particles and material activation on the SAA exits. During quiet orbital phases, the background remains close to Monte-Carlo predictions. The time periods when the count rates are null correspond either to XOFF period or passage through the deep SAA -- see Section~\ref{sect:ugts}.}
        \label{fig_bkgvsmc}
    \end{figure}

The in-flight background level is 2700 -- 2900 cts s$^{-1}$ in the 4--120 keV range when the Earth is outside the ECL FoV and particle contributions are negligible. This in-flight measure is consistent with our GEANT4-based simulation results (\citealt{Mate19, Bouchet26}). The background simulator performs dynamic modeling that accounts for the evolution of the orbital parameters and the Earth motion across the FoV (Fig.~\ref{fig_bkgvsmc}). When the FoV is fully occulted by the Earth, the background is then dominated by the reflected CXB and Earth albedo emission (\citealt{Cordier08}).

\subsubsection{SAA contours}
\label{sect:SAA}

Because {\it SVOM} is in a low Earth orbit (LEO) with an inclination of 29°, ECL crosses the deep SAA 5–6 times per day. In this region, trapped protons and electrons generate high count rates in the instruments, requiring specific operational management (see Section~\ref{sect:ugts}, \citealt{Claret25}).
The SAA contours depend on the instrument and energy band and are defined using count rates from different ECL event types. Although they evolve with terrestrial and solar magnetic activity, a quasi-static base contour can be established.
Before launch, contours were modeled with NASA AE8/AP8 particle fluxes and ECL Monte-Carlo simulation. Results were validated against Demeter electron data and {\it Swift}/BAT contours, which also makes use of similar detectors (CZT\,\footnote{Alloy made of Cd, Te and Zn}) and operates in LEO.

\begin{figure}
        \centering
        \includegraphics[width=1.1\linewidth, angle=0]{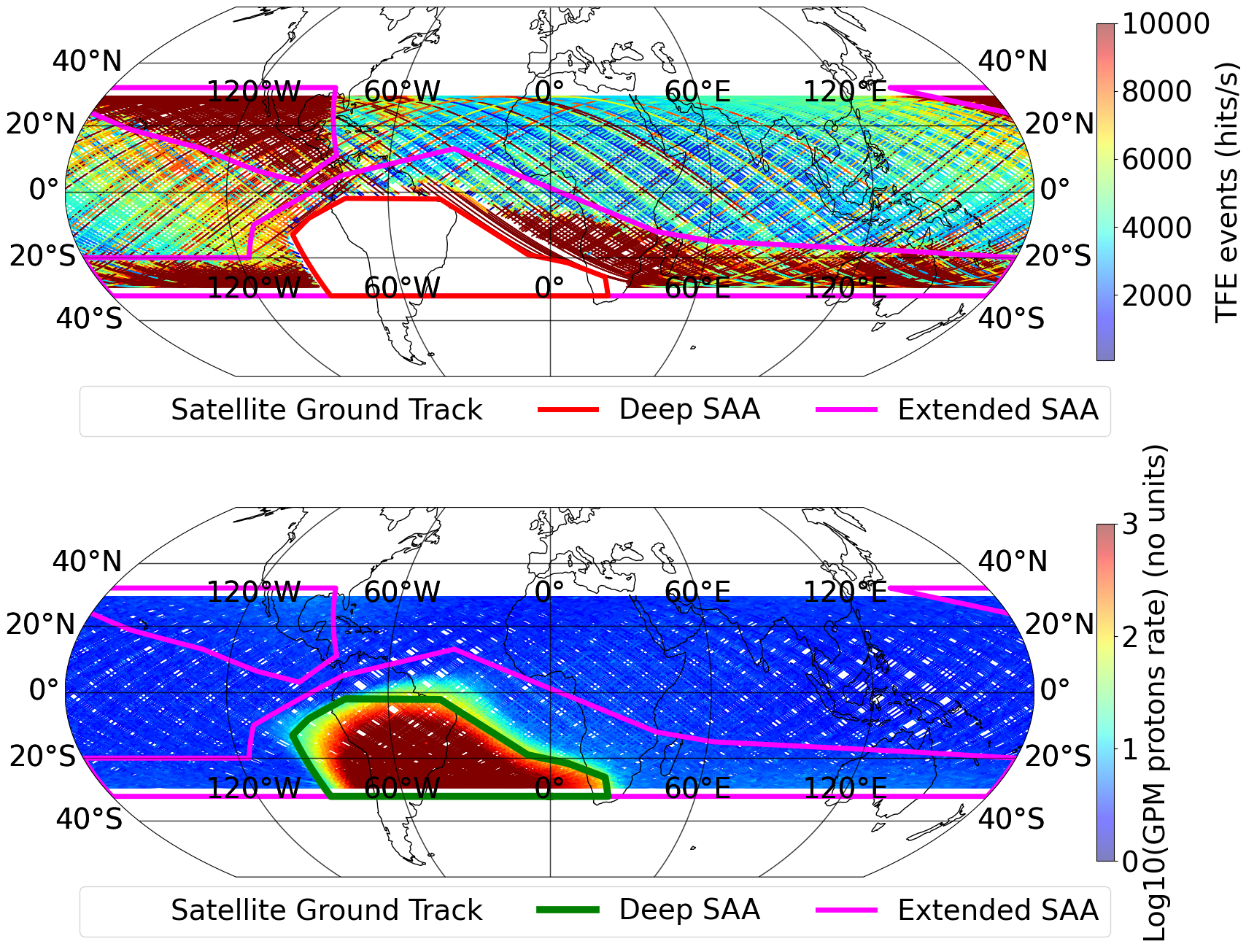}       
        \caption{(Top) Deep and extended SAA contours based on the TFE count rates measured by ECL. (Bottom) SAA contour estimation based on proton count rates measured by the GRM particle monitor (GPM) onboard {\it SVOM}. Note the very good consistency between the two approaches.}
        \label{fig_SAA}
    \end{figure}

After launch, SAA contours were refined using in-flight data (e.g. TFE count rates). With 15 orbits per day and continuous measurement of proton counts from the GRM Particle Monitor (GPM), the contours are derived from the evolution of TFE and GPM count rates along the orbit and geographic coordinates.
High-energy protons produces a trailing background through activation when the satellite exits the deep SAA. Figure~\ref{fig_SAA} shows the deep (in orange) and extended (in green) SAA contours currently used onboard ECL, overlaid on TFE and GPM count rates along the orbit. When exiting the deep SAA, an additional trailing background induced by activation is seen. Given the high count rates produced, the XOFF mechanism is often activated up to the point when this background component decreases significantly (see \citealt{Claret25} and the TFE count rates on the eastern part of the deep SAA in Fig.~\ref{fig_SAA}).

\subsection{Spectral performances}
\label{sect:spec_perfo}

\subsubsection{Energy scale \& low-energy threshold}
\label{sect:nrj}

{\it Energy scale} -- The energy scale of events encoded in channels ($C_{\rm evt}$) is reconstructed in keV via the linear scale: $E_{\rm evt} = C_{\rm evt} \times G + O$, with $E_{\rm evt}$ the event energy derived with the gain $G$ in keV/channel and the offset $O$ in keV. Each of the 6400 pixels has its own relation that has been calibrated on-ground making use of different X-ray lines from 6 keV to 122 keV (\citealt{Godet22, Godet25}). The gain and offset information of each pixel of the detection plane is stored as a matrix in our calibration database (CALDB).  For use by the onboard trigger software, the above gain--offset relation is converted into onboard configuration tables containing the EBANDS limits in channel for each of the 6400 pixels (see Section~\ref{sect:data} and \citealt{Schanne25}).

During flight, the gain--offset relation of each pixel is susceptible to evolve, which therefore requires monitoring.  Using fluorescence K-shell lines obtained during regular observations, one can monitor on the ground any significant channel shift of the line centroids, and adjust the gain--offset relation as needed. The most prominent lines for this purpose are the Pb lines from 70 to 90 keV, the Cu line at $\sim 8$ keV, and the Cd and Te lines from 23 to 31 keV.  In practice, determining the instrumental channel of the line centroid requires long cumulated exposures, as well as low background.  This is particularly true for the low-energy lines, which are dominated by the CXB.  Limiting the effects of the CXB can be achieved when the Earth blocks the camera FoV, but this reduces the available exposure time.

 \begin{figure}
        \centering
        \includegraphics[width=0.9\linewidth]{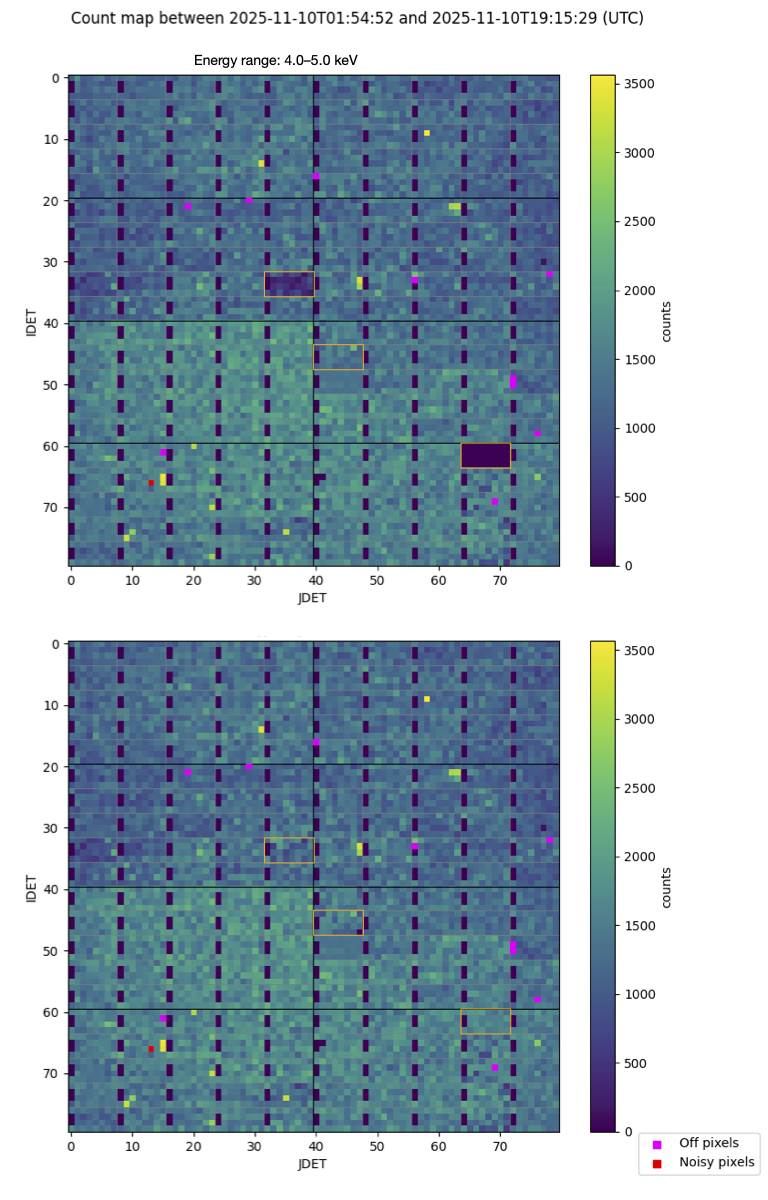}
        \caption{Background shadowgrams in the 4 -- 5 keV band before (top) and after (bottom) the energy scale correction of the three modules surrounded in white.}
        \label{fig:nrj_scale}
    \end{figure}
    
During the commissioning phase following a reboot of ECL, a module (i.e. 32 pixels) had its energy scale drastically modified, shifting events to higher energies by $> 1$ keV.  For the onboard source trigger software, this caused multiple false triggers due to the resulting deficit or excess of events in three of the four ESTRIP intervals (see Section~\ref{sect:data}. Masking this module in the onboard deconvolution process of the trigger algorithm reduced the occurrence of false triggers, but the energy scale of the entire module had to be corrected. 
Figure~\ref{fig:nrj_scale} shows background shadowgrams in the 4 -- 5 keV band before (top) and after (bottom) the energy scale correction of this module. The corrected EBANDS table also includes corrections for three other modules, which had minor shifts of their energy scale compared to data collected prior to the launch. These three modules were also masked out for the onboard deconvolution process. The correction of the EBANDS table allowed to re-include these modules and  retrieve the full effective area of ECL.

{\it Low-energy threshold} -- As discussed in \cite{Godet22}, we calibrated before launch the linear SBN\,\footnote{SBN stands for digital low energy threshold in English.} -- $E_\mathrm{th}$ relation: $E_\mathrm{th} = \alpha \times \mathrm{SBN} + \beta$ for the 6400 pixels of the detection plane. $E_\mathrm{th}$ is the pixel low-energy threshold in keV while SBN is an integer from 0 to 62\,\footnote{SBN = 62 roughly corresponds to a low-energy threshold of 14 keV.} allowing to choose the value of the ASIC low-level discriminator. A value of SBN = 63 indicates that the pixel is no longer read out by the electronic chain (i.e. it is disabled). We set the pixel $E_\mathrm{th}$ value at 3.8 keV for 93.75\% of the pixels and at 7 keV for the rest of them. The reason for the 7 keV threshold was to mitigate some cross-talk effects induced by two adjacent pixels on each XRDPIX module (see Fig.\,4 in \citealt{Godet25}). On-ground long-term thermal-vacuum tests on the ECL in-flight model simulating expected temperature variations of the detectors along the {\it SVOM} orbit ($\pm 3$°C) around a nominal value demonstrated that the pixel low-energy threshold is stable within one increment of the SBN-value (corresponding to $\sim 0.24$ keV in average) -- see Fig. 19 in \cite{Godet22}. Using in-flight background data, we confirm the good stability over time and with temperature variation along the orbit of the low-energy threshold (see Fig.\,3 in \citealt{Godet25}).

\subsubsection{Spectral response}
\label{sect:spectral_response} 

To evaluate the spectral performance of ECL, we focus on Crab observations, which serves as a standard candle in X-ray astronomy. During the commissioning phase, 36 off-axis observations were performed to verify that the spectrum is properly reconstructed by the ECL Pipeline (ECPI -- \citealt{Goldwurm25}) for different azimuthal angles $\phi$ and at an off-axis angle of $\theta = 30^\circ$. Within a single observation, data can be either not occulted by the Earth (No Earth Occultation, NEO) or partially occulted (Partially Earth Occultation, PEO). The stacked combination of these data for a given observation is referred to as NPO.

\begin{figure}
        \centering
        \includegraphics[width=0.95\linewidth]{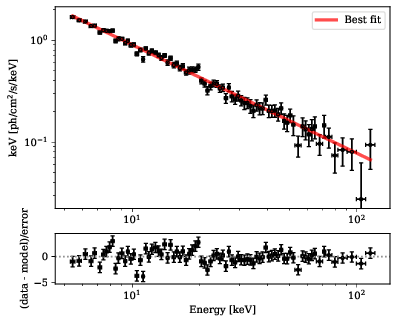}
        \caption{Crab spectrum obtained for NPO data (i.e. stacked NEO and PEO data) from the observation on October 23 2024, with a total exposure time of 1307\,s, an azimuthal angle of $\phi = 330^\circ$, and an off-axis angle of $\theta = 30^\circ$.}
        \label{fig:crab_spectrum}
    \end{figure}

For the present analysis, we used ECPI version 1.18.2. Spectra were extracted into 96 energy bins. The modules presenting abnormal energy scale (see Section~\ref{sect:nrj}) were excluded from the data analysis.
Details on the ECL spectral extraction can be found in the appendix of \citealt{Cordier25a}. The spectra were fitted with a powerlaw model in \textsc{xspec} version 12.14.1, without the addition of systematic uncertainties. Figure~\ref{fig:crab_spectrum} presents an example of the reconstructed spectrum. It corresponds to a stacked spectrum from NEO and PEO data collected during the observation of October 23, 2024, with a total exposure time of 1307\,s, for an azimuthal angle of $\phi = 330^\circ$ and an off-axis angle of $\theta = 30^\circ$.

\begin{figure}
        %\centering
        \includegraphics[width=1.05\linewidth]{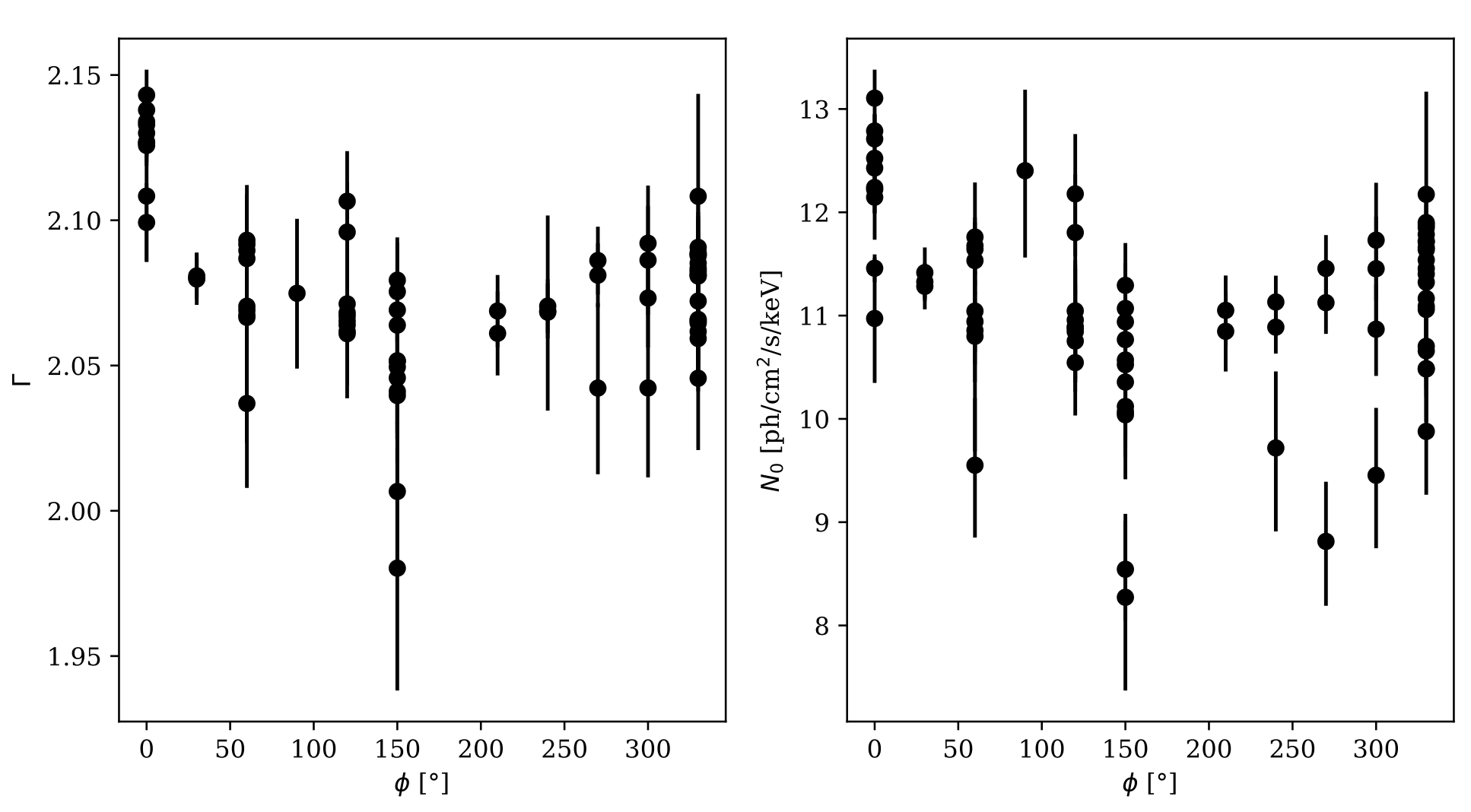}
        \caption{Evolution of the Crab spectral parameters $\Gamma$ and $N_0$ as a function of the azimuthal angle $\phi$. Error bars correspond to 90\% c.l.}
        \label{fig:evolution_phi_param}
    \end{figure}

The evolution of the photon index $\Gamma$ and the normalization $N_0$ at 1 keV as a function of the azimuthal angle $\phi$ is shown in Fig.~\ref{fig:evolution_phi_param}.
The corresponding mean values for $\Gamma$, $N_0$, and the 4--120\,keV flux are 
$\overline{\Gamma} = 2.08 \pm 0.02$, $\overline{N_0} = 11.3 \pm 0.8$\,ph\,cm$^{-2}$\,s$^{-1}$\,keV$^{-1}$, and $\overline{F}_{\mathrm{4-120\,keV}} = (4.8 \pm 0.2) \times 10^8$ ergs cm$^{-2}$\,s$^{-1}$. The parameters obtained for NEO, PEO, and NPO observations are reported in Table~\ref{tab:parameters}. 

Overall, these first results making use of the spectral response files computed prior to the launch (\citealt{Bouchet26}) are very good. Indeed, the spectral parameters are mostly consistent across observations, regardless of the azimuthal angle or the Earth occultation, even if there are some small discrepancies within the spectral parameters in some cases. Some background correction issues in some cases might account for a part of this dispersion. Note that some residuals around 14 -- 25 keV may be seen in spectra with high statistics like for the Crab. We advise -- when it is the case -- to add 5\% systematics in the 14 -- 25 keV band. The ECL team continues working on these minor issues to further consolidate the results obtained on the spectral response.

\begin{figure}
        \centering
       \includegraphics[width=0.95\linewidth]{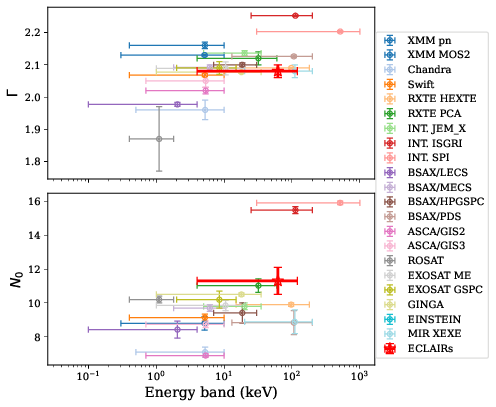}
        \caption{Spectral parameters of the Crab for various X-ray missions (adapted from \citealt{Kirsch2005}).}
        \label{fig:all_missions}
    \end{figure}

 Figure~\ref{fig:all_missions} shows the spectral parameters obtained with various X-ray missions (\citealt{Kirsch2005}), together with those measured by ECL. The spectral parameters derived with ECL are consistent with those obtained from other X-ray missions, in particular with instruments operating in a similar energy range, demonstrating the very good spectral performance of ECL. 

    \begin{table}[h!]
        \centering
        \caption{Mean spectral parameters of the Crab obtained for NEO, PEO, and NPO (stacked NEO and PEO) observations. The 1$\sigma$ dispersion on the spectral parameters derived at different angles is also quoted.}
        \label{tab:parameters}
         \begin{tabular}{lccc}
        \hline
        Parameter & NEO & PEO & NPO \\
        \hline
        $\Gamma$ & $2.09 \pm 0.02$ & $2.06 \pm 0.03$ & $2.07 \pm 0.02$ \\
        $N_0$ (ph\,cm$^{-2}$s$^{-1}$keV$^{-1}$) & $11.6 \pm 0.6$ & $10.8 \pm 1.0$ & $11.3 \pm 0.7$ \\
        Flux (4--120\,keV) & $4.8 \pm 0.2$ & $4.8 \pm 0.2$ & $4.8 \pm 0.1$ \\
         ($\times 10^{-8}$ ergs/s/cm$^2$) &  &  &  \\
        \hline
        \end{tabular}
    \end{table}

\subsubsection{Effective area}
\label{sect:EA} 

\begin{figure}
        \centering
        \includegraphics[width=\linewidth]{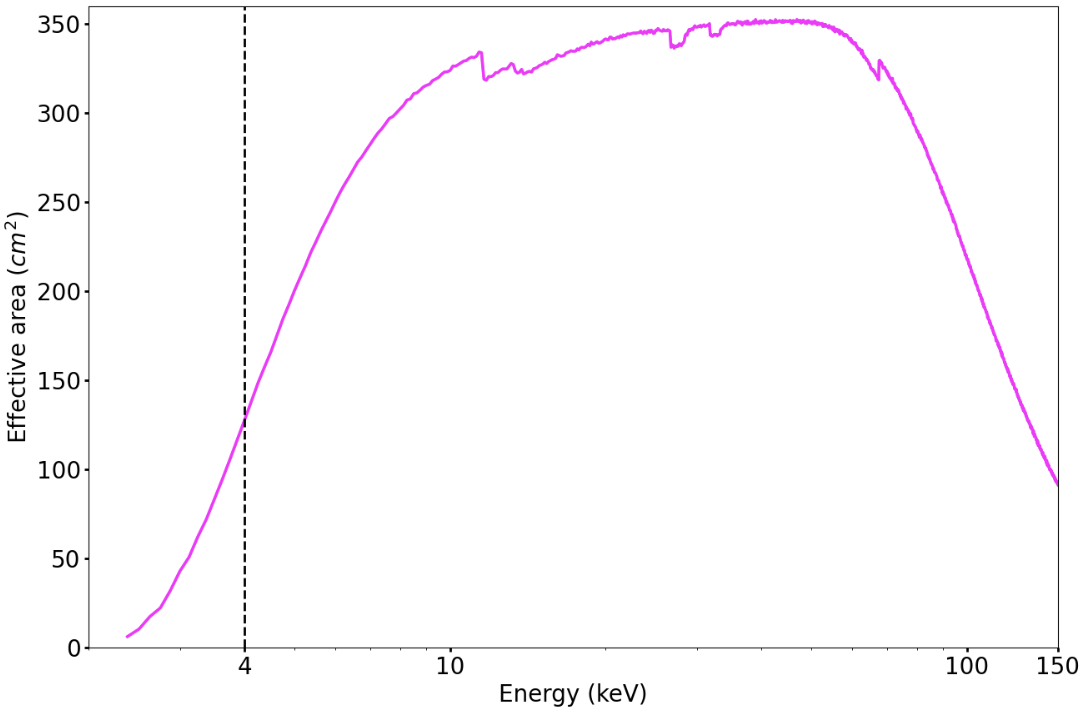}
        \caption{ECL effective area (on-axis) for single events as function of photon energy. Below 20 keV, the Pt L-shell absorption edges are clearly visible. The two dips at 26.7 and 31.8 keV are due to Cd and Te fluorescence lines being considered as multiple events. The discontinuity at 67.4 keV is due to K-shell edge from Ta.}
        \label{fig:EA}
    \end{figure}

During the on-ground calibration of the ECL in-flight model, we performed some measurements of the camera effective area (EA) as a function of energy using X-ray lines from radioactive sources (from 6 keV to 122 keV) as well as fluorescence X-ray lines produced by an X-ray generator for photon energies below 20 keV to have a finer view on the evolution of EA at low energies. Comparisons between the measured and GEANT4-based computed EA have been made showing agreement within $\pm 5\%$ across the ECL energy band; which is fully compliant with the ECL science requirements (see Fig. 15 in \cite{Godet22} -- see also \cite{Bouchet26} for more detail on the EA GEANT4-based computation and comparisons to the data). Figure~\ref{fig:EA} shows the on-axis ECL effective area in the 4 -- 150 keV energy range for single events after the deconvolution process. Note that photons not modulated through the coded mask are lost in the deconvolution process.

\subsection{Imaging performances}
\label{sect:imaging}

{\it Localization bias matrix} -- 
An estimation of the localization biases was performed using the imaging procedure of the ECPI (\citealt{Goldwurm25}). Using observations of the Crab and the Cygnus region (Cyg X-1, Cyg X-2, and Cyg X-3) taken during the first year of operation of ECL, we compare the reconstructed directions of the sources to the actual ones. These offsets were minimized using a least-squares minimization method, taking into account four parameters: the three misalignment angles between the platform attitude and the telescope pointing, and the mask height from the detector plane. 
A significant discrepancy of +0.11 cm was found between the real mask height and the nominal one of 45.77 cm. Correcting for this shift significantly improves the localization of the sources. On the other hand, the misalignment was found to be marginal, with an on-axis rotation (X axis) of $\approx 0.3’$, and image axes Y and Z rotations of $\approx 0.7’$ and $\approx 0.9’$, respectively.  

\medskip

{\it Point Source Location Error} -- 
The PSLE was estimated after correction of the localization biases (\citealt{Goldwurm25}). This error depends on the SNR of the source, with a theoretical relation at a 90\% c.l. in an ideal detector of $\textrm{PSLE}_{90\%} = 2.25 ~\mathrm{sky~pixels} / \mathrm{SNR}$ (\citealt{Goldwurm22}). 
The in-flight relation was estimated using all available good-quality sky images of the Crab and the Cygnus region (see \citealt{Goldwurm25} for more detail).  
The positioning error at a 90\% c.l. follows the trend given by $\textrm{PSLE}_{90\%} = (3.35 \pm 0.03)/\mathrm{SNR} + (3.5 \pm 0.5) \times 10^{-2}$, in units of sky pixels. For reference, 1 sky pixel on axis covers 34’. This trend exceeds the one predicted for an ideal detector and shows a plateau at high SNR. This suggests that some systematic effects could be improved with future calibration and/or data analysis software improvements. 

  \begin{figure}
        \centering
        \includegraphics[width=\linewidth]{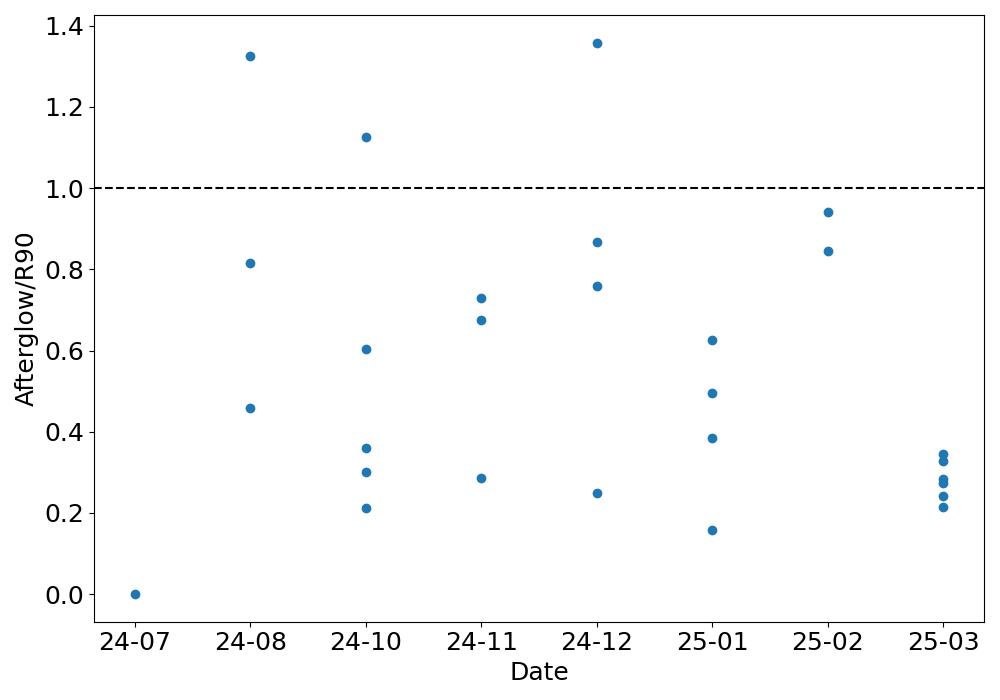}
        \caption{Angular distance from the ECL GRB positions to the afterglow ones normalized by the onboard ECL 90\% c.l. PSLE up to March 31, 2025.}
        \label{fig:position}
    \end{figure}

Onboard, the currently used 90\% c.l. PSLE is given by: $\textrm{PSLE}_{90\%} = \sqrt{\left(\frac{77'}{\mathrm{SNR}}\right)^2 + \mathrm{syst^2}}$ with a systematic error of $\mathrm{syst} = 2'$. Even if the onboard $\textrm{PSLE}_{90\%}$ relation could be further improved, in most cases the GRB afterglows were found within this PSLE relation as shown in Figure~\ref{fig:position}.

\subsection{Timing performances}
\label{sect:timing}

An estimation of the ECL capabilities for timing analyses was performed using Crab pulsar observations. This source, with a spin frequency of $\sim$ 29.5 Hz, is regularly monitored by radio facilities. Its frequency and frequency derivative were retrieved from public data from the Jodrell Bank Crab Pulsar Monthly Ephemeris\footnote{\url{www.jb.man.ac.uk/~pulsar/crab.html}}. 
A single-pointing observation of 7 hours, including the Crab in the FoV, was taken on February 6, 2025. The pulsar frequency was computed by epoch-folding with $Z^2_n$ maximization (\citealt{Lyne_1993}), and the difference from the radio frequency was found to be $(0.5 \pm 1.0) \times 10^{-5}$ Hz, and fully compatible within a relative uncertainty of $\Delta\nu/\nu = 1.7 \times 10 ^{-7}$. On the same observation, no variation of the frequency could be detected, and an upper limit was computed at $\dot\nu < 3 \times 10^{-9}$ Hz/s, which is compatible with the actual variation measured in radio of $\dot\nu = 3 \times 10^{-10}$ Hz/s.

The absolute timing capabilities were estimated similarly, by comparing the phase of the primary pulsar peak to the one given by the radio ephemeris. This computation was performed on a set of 13 observations containing the Crab in the FoV. While the frequencies remained compatible with the ones measured in radio, a constant systematic drift of $\sim 150\,\mu$s per day of the phase was observed. This effect is still under investigation at the time of writing, but may be a consequence of a slight dragging of the internal clock used by the instrument. The impact on the frequency measurement is a relative deviation of the order of $10^{-9}$, which should be negligible for most timing analyses. In any case, a re-calibration using radio measurements as references is sufficient to recover an adequate absolute timing.

\section{Science performances}
\label{sect:science}

This section serves as a conclusion for the series of ECL papers. They outline the excellent performance of the ECL instrument (\citealt{Lachaud25, Godet25, Schanne25}) and the good maturity of the data analysis pipelines (\citealt{Piron25, Goldwurm25}). Here, we outline the uniqueness of ECL to perform cutting edge science by discussing some science results for the GRB and observatory science programs obtained before May 31, 2025.

\subsection{Triggers and duty cycle}
\label{sect:trigger}

Thanks to the very good performance and behavior of the camera, we were able to activate the onboard trigger algorithms as early as on July 12 2024, resulting in detecting the first ECL GRB on July 13 2024 (GRB\,240713A) (\citealt{Schanne24})! This first burst was later confirmed as a subthreshold event by Fermi/GBM.

  \begin{figure}
        \centering
       \hspace{-10mm} \includegraphics[width=1.1\linewidth]{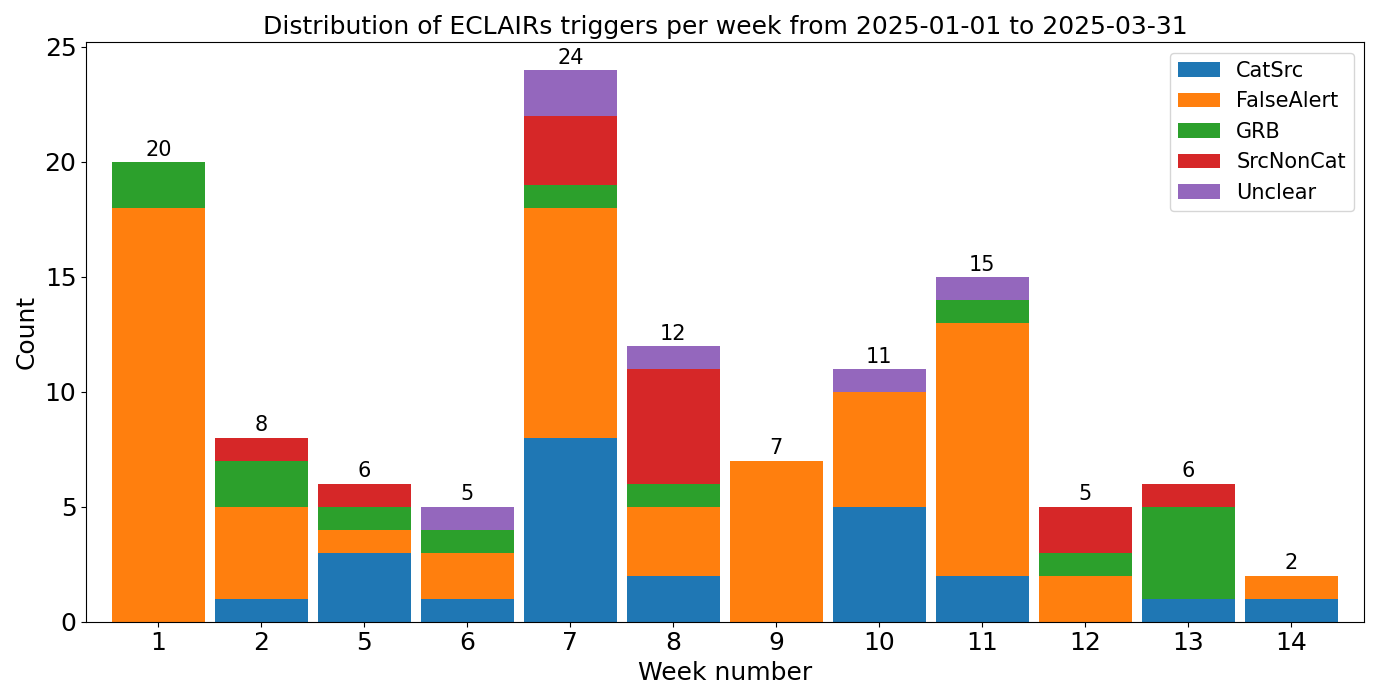}
        \caption{ECL triggers from January 1, 2025 up to March 31, 2025. CatSrc correspond to sources present in the onboard source catalog, while SrcNonCat refer to known sources not included in the onboard source catalog. False Alerts could be due to multiple factors (e.g. coding noise from bright sources not properly corrected in the shadowgrams, particle background, etc.)}
        \label{fig:trigger}
    \end{figure}

Figure~\ref{fig:trigger} shows the number of ECL triggers per month as well as the breakdown of their nature. The false triggers have multiple origins: coding noise from bright X-ray sources (e.g. Sco X-1, the Crab) not being properly corrected when within the ECL FoV, instrumental noise (e.g. noisy modules or columns of modules -- see \citealt{Godet25}), Earth magnetosphere phenomena (e.g. high particle background, possible X-ray auroras). In most cases, we learned how to limit the number of false triggers in such a way that they did not have any significant damaging impacts on the mission and instrument systems.

\begin{figure}
        \centering
        \includegraphics[width=1.12\linewidth]{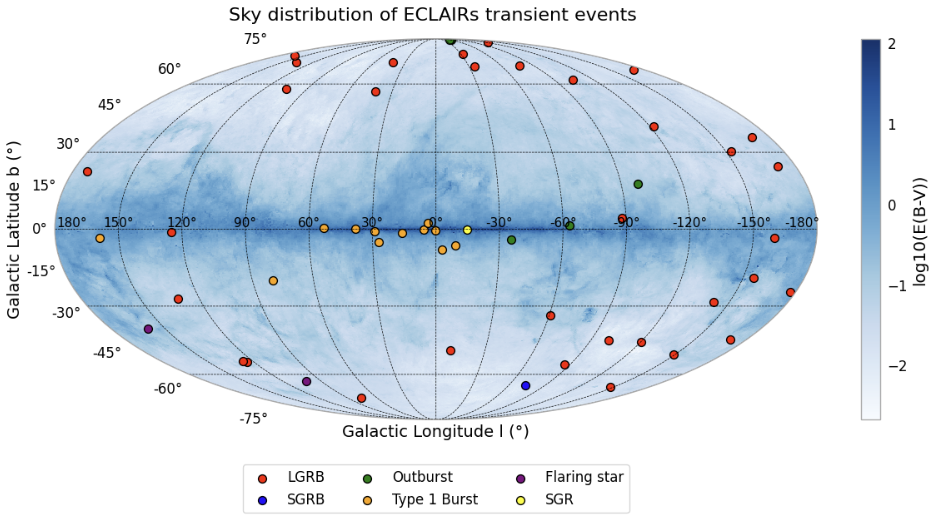}
        \caption{Sky localization of the 32 GRBs detected by ECL (29 onboard $+$ 3 on-ground) and Galactic sources producing Type-1 X-ray bursts (mostly on-ground) up to March 31, 2025. The color scale corresponds to the Galactic extinction in different sky directions.}
        \label{fig:sky}
    \end{figure}

Up to March 31 2025, ECL detected onboard 29 GRBs of all types: i) One short GRB with an extended tail (GRB 240821A); ii) Several X-ray flashes like GRB 241001A (see Section~\ref{sect:GRB}) and X-ray rich GRBs; iii) GRB 250314A, a burst at $z\sim 7.3$ associated to a possible type Ib/c supernova (\citealt{Cordier25, Levan25}); iv) Two very long (a few minutes) GRBs like GRB 241217A  -- see Figure~\ref{fig:sky} and \cite{Daigne25}.  
From these 29 detected GRBs, 6 (21\%) have their best SNR alert in the 5 -- 8 keV ESTRIP band.

The GRB detection rate is strongly dependent on the observing duty cycle of the instrument as seen in Section~\ref{sect:ugts}. In addition, before December 2024, ECL was not operating nominally because of the various commissioning activities. This reduced the instrument observing duty cycle. Since December 2024, ECL operates nominally increasing the duty cycle. All in all we estimate the onboard detection rate to be around 50--60 GRBs per year; which is commensurate with estimates done prior to {\it SVOM} launch.

In addition to the onboard detected GRBs, on-ground searches with the offline trigger pipeline (OFTG -- \citealt{Brunet26}) making use of the X-band event-by-event data nicely complemented the onboard source trigger software by detecting 3 long GRBs. A description of the OFTG main functionalities is given in \cite{Coleiro25}.
By using a different configuration and phasing from those used in the onboard source trigger software, the OFTG allows to probe sources that cannot be detected onboard (e.g. the on-ground detection of EP 241208a -- \citealt{Atteia24}). It already allowed to improve the onboard detection efficiency by fine tuning some energy strips.

\subsection{GRB science}
\label{sect:GRB}

As mentioned in Section~\ref{sect:trigger}, ECL detected onboard 29 GRBs up to March 31 2025,  of which 12 have measured redshifts spanning a broad range of redshifts, from z = 0.238 to 7.3 -- \cite{Daigne25}. VT afterglow follow-ups of ECL GRBs appear to show unique lightcurves with features not so commonly seen in optical from GRBs detected by other GRB missions like {\it Fermi} and {\it Swift} (e.g. transitions from the prompt-to-afterglow phase, late emission peaks of the optical afterglow, complex temporal breaks at early and late times).

  \begin{figure}
        \centering
        \includegraphics[width=\linewidth]{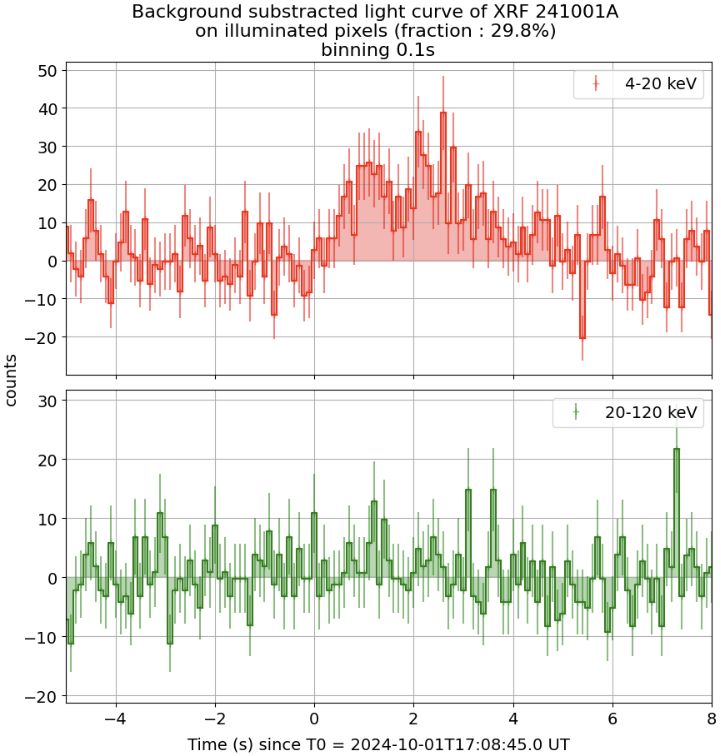}
        \caption{Background subtracted ECL lightcurves of GRB\,241001A in the 4 -- 20 keV and 20 -- 120 keV bands. This long ($\sim 3$ s) burst can be classified as an XRF given that most of the emission is seen below 20 keV and its spectrum could be fitted with a black-body model with $kT \sim 1.7$ keV (\citealt{Schneider26}).}
        \label{fig:GRB241001A}
    \end{figure}

Thanks to its 4 keV low energy threshold, ECL is able to detect soft ($< 20$ keV) GRBs with $E_\mathrm{peak}$ values $< 5$\,keV or with thermal-like spectral shapes like GRB 241001A (z = 0.57) associated with a type Ib/c supernova detected by JWST (see Fig.~\ref{fig:GRB241001A}). That latter one  would not have been detected by {\it Swift}/BAT being too faint and too soft (\citealt{Schneider26}). In addition, some of these soft GRBs -- even if getting prompt follow-ups by VT and/or optical robotic telescopes providing early deep upper limits -- somehow challenge the present follow-up strategy. Thus, this prevents deciphering their true nature. However, these bursts are of potential great interest for the GRB community since some of them may be high-z GRBs.

\begin{figure}
        \centering
        \includegraphics[width=\linewidth]{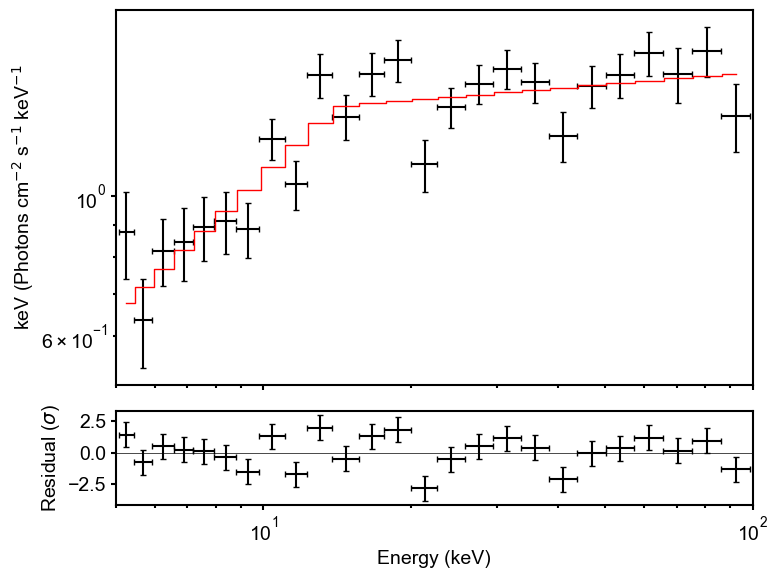}
        \caption{Spectrum of the GRB 250506A prompt emission and its best-fit broken powerlaw model (solid line). The break energy is around 14 keV. Note that {\it Fermi}/GBM measured an E$_\mathrm{peak}$-value around 280 keV for this burst. }
        \label{fig:GRB250506A}
    \end{figure}

In addition, standalone ECL spectral analyses could reveal the presence of low-energy spectral features like low-energy breaks as seen in GRB 250506A (see Fig.~\ref{fig:GRB250506A}) with a break energy around 14 keV (\citealt{GCN_GRB250506A}). This spectral break is significant at 5.1 $\sigma$ level. {\it Fermi}/GBM measured an E$_\mathrm{peak}$ value around 280 keV for this burst (\citealt{GCNFermi}). Such low-energy spectral features are of great interest since they could help better understand the physical processes producing the GRB prompt emission. The ECL team started looking for such low-energy spectral features/breaks by performing broadband spectral analysis including data at energies lower than 4 keV with EP/WXT and at higher energies with GRM and/or {\it Fermi}/GBM.

\subsection{Observatory science}
\label{sect:nonGRB}

In addition to GRBs, ECL capabilities open numerous observatory science studies, either on short non-GRBs transients (seconds to minutes) detected onboard and/or with the OFTG (\citealt{Brunet26}) or longer timescales (e.g. monitoring of sources over several weeks/months) with the Quick Look Analysis tools (\citealt{Goldwurm25, Coleiro25}).

\begin{figure}
        \centering
        \includegraphics[width=\linewidth]{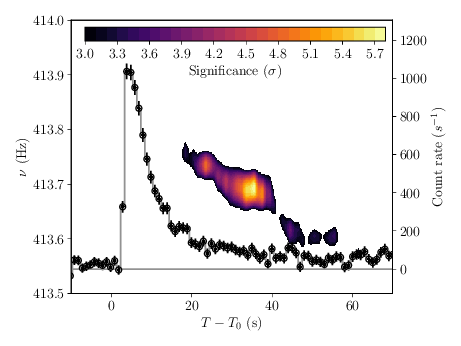}
        \caption{1\,s binned lightcurve in the 4 -- 40 keV band of a Type-I X-ray burst in the NS-LMXB 4U~0614+091. The persistent emission is subtracted. The color scale shows the oscillation frequency found during the burst and its downshift. Image from \cite{Lestum26}.}
        \label{fig:4U}
    \end{figure}

Among the short transients are the thermonuclear X-ray bursts, indicating nuclear burning at the surfaces of accreting neutron stars (NS) in low-mass X-ray binaries (LMXBs, \citealt{GK21}). The wide ECL FoV as well as its sensitivity down to 4~keV, has made possible the detection of these X-ray bursts, generally serendipitously, when looking at or near the Galactic plane (see Fig.~\ref{fig:sky}). Most of these bursts have been detected by the OFTG because the onboard source trigger software has been optimized to detect extragalactic transients like GRBs and some bursts are just too faint. Note that the onboard trigger software detected several bursts/outbursts from Galactic sources included or not in the onboard source catalog (see Fig.~\ref{fig:trigger}). Time-resolved spectral analyses can then be performed over the duration of the burst to study the temperature evolution, and even determine whether the burst belong to the so-called "photospheric radius expansion" category (\citealt{GK21}).   
A handful of those have been detected by ECL and published through Astronomer's Telegram (e.g. \citealt{ATEL1, ATEL2}).  Another interesting result is the discovery of burst oscillations with a decreasing frequency during a Type I X-ray burst of the NS-LMXB 4U~0614+091 (\citealt{Lestum26}). While the oscillation average frequency ($f\sim 413.674$~Hz) corresponds to the NS spin, the downward frequency drift may be interpreted as the effects of the orbital motion, which allows us to put constrain on the orbital period (see Fig.~\ref{fig:4U}). As ECL has detected $>50$ X-ray bursts after 8 months of operation, this wealth of information on NS-LMXBs will be compiled into a catalog of {\it SVOM} X-ray bursts.  
In addition to astrophysical search for short timescale transient, ECL thanks to regular transits of the Earth within its FoV also allows searching for more local phenomena like terrestrial gamma-ray flashes, ms-scale transients associated to particle acceleration during thunderstorms (\citealt{Tavani11}) in synergy with GRM.

On longer timescales, ECL can monitor transient activity associated with stellar flares, AGN, X-ray binaries, etc -- see \cite{Coleiro25}.  Regarding stellar flares that are thought to find their origin in magnetic reconnection (e.g. \citealt{Shulyak17}), these events release $10^{33}$ -- $10^{39}$~ergs of energy within tens to hundreds of minutes.  The ECL transients usually require other wavelengths detections (with VT and/or our ground robotic telescopes) to confirm the flaring activity. This was the case of the stellar flare from the RS-CVn star HD~22468 (\citealt{Wang26}). The ECL detection and subsequent observations over $\sim2.6$~hours permits studies of the spectral evolution of this flare (e.g., plasma temperature).

\begin{acknowledgements}
The Space-based multi-band astronomical Variable Objects Monitor (SVOM) is a joint Chinese-French mission led by the Chinese National Space Administration (CNSA), the French Space Agency (CNES), and the Chinese Academy of Sciences (CAS). We gratefully acknowledge the unwavering support of NSSC, IAMCAS, XIOPM, NAOC, IHEP, CNES, CEA and CNRS. We thank all the persons (engineers, scientists, Ph-D students, postdocs, sub-contractors, etc.) who have worked hard during more than 15 years to build a fantastic instrument. The ECL team dedicates this paper to Pierre Mandrou, a missed friend and colleague.

\end{acknowledgements}

\label{lastpage}


\begin{thebibliography}{99}
%% you can type \apj for ApJ, \aap for A&A, \apss for Ap&SS, etc. Please consult
%% the macro chjaa.cls. You can also find them in aasguide.tex (AASTeX for ApJ, AJ, PASP)
%% Please follow the format of ChJAA's reference list

\bibitem[Ajello et al. (2008)]{Ajello08} Ajello, M., Greiner, J., Sato, G. et al., 2008, \apj, 689, 666, \href{https://ui.adsabs.harvard.edu/link_gateway/2008ApJ...689..666A/doi:10.1086/592595}{10.1086/592595}  

\bibitem[Atteia et al. (2024)]{Atteia24} Atteia, J.-L., Bouchet, L., Brunet, M. et al., 2024, GRB Coordinates Network, Circular Service, No. 38557

  \bibitem[Bouchet et al. (2026)]{Bouchet26} Bouchet, L., Godet, O., Atteia, J.-L. et al., 2026, in prep. 

  \bibitem[Brunet et al. (2026)]{Brunet26} Brunet, M., Yang, H., Bouchet, L. et al., 2026, in prep. 
  
  \bibitem[Brunet et al. (2025a)]{ATEL1} Brunet, M., A. Coleiro, Guillot, S., Rodriguez, J., Cangemi, F., Zhang, L., 2025, Astronomer's Telegram, 17251

 \bibitem[Brunet et al. (2025b)]{ATEL2} Brunet, M., Cangemi, F., Guillot, S., Coleiro, A., Zhang, L., 2025, Astronomer's Telegram, 17385 

 \bibitem[Churazov et al. (2008)]{Churazov08} Churazov, E., Sazonov, S., Sunyaev, R. and Revnivtsev, M. 2008, MNRAS, 385, 719, \href{https://ui.adsabs.harvard.edu/link_gateway/2008MNRAS.385..719C/doi:10.1111/j.1365-2966.2008.12918.x}{10.1111/j.1365-2966.2008.12918.x}
 
  \bibitem[Churazov et al. (2007)]{Churazov07} Churazov, E., Sunyaev, R., Revnivtsev, M. et al., 2007, \aap, 467, 529, \href{https://ui.adsabs.harvard.edu/link_gateway/2007A&A...467..529C/doi:10.1051/0004-6361:20066230}{10.1051/0004-6361:20066230}

  \bibitem[Claret et al. (2025a)]{Claret25} Claret, A., Bouchet, L., Godet, O. et al., 2026, IEEE Transactions on Nuclear Science, submitted
  
  \bibitem[Coleiro et al. (2026)]{Coleiro25} Coleiro, A., Tao, L., Cangemi, F. et al., 2026, RAA, this issue  
  
  \bibitem[Cordier et al. (2026)]{Cordier25} Cordier, B., Wei, J. Y., Zhang, S.N. et al., 2026, RAA, this issue

  \bibitem[Cordier et al. (2025)]{Cordier25a} Cordier, B., Wei, J. Y., Tanvir, N. R. et al., 2025, A\&A Letter, 704, L7, arXiv:2507.18783, \href{https://ui.adsabs.harvard.edu/link_gateway/2008AIPC.1000..585C/doi:10.1063/1.2943538}{10.1063/1.2943538}


   \bibitem[Cordier et al. (2008)]{Cordier08} Cordier, B., Desclaux, F., Foliard, J. and Schanne, S., 2008, GAMMA-RAY BURSTS 2007: Proceedings of the Santa Fe Conference. AIP Conference Proceedings, 1000, 585, arXiv:0807.0739, \href{https://ui.adsabs.harvard.edu/link_gateway/2008AIPC.1000..585C/doi:10.1063/1.2943538}{10.1063/1.2943538}

  \bibitem[Dagoneau et al. (2021)]{Dagoneau21} Dagoneau, N., Schanne, S., Rodriguez, J., Atteia, J.-L. and Cordier, B., 2021, \aap, 645, A18, \href{https://ui.adsabs.harvard.edu/link_gateway/2021A&A...645A..18D/doi:10.1051/0004-6361/202038995}{10.1051/0004-6361/202038995}

  \bibitem[Daigne et al. (2026)]{Daigne25} Daigne, F., Turpin, D., Atteia, J.-L. et al., 2025, RAA, this issue


   \bibitem[Galloway \& Keek (2021)]{GK21} Galloway, D. K. and Keek, L., 2021, Astrophysics and Space Science Library, 461, \href{https://ui.adsabs.harvard.edu/link_gateway/2021ASSL..461..209G/doi:10.1007/978-3-662-62110-3_5}{10.1007/978-3-662-62110-35}
 
  \bibitem[Gevin et al. (2009)]{Gevin09} Gevin, O., Baron, P., Coppolani, X. et al., 2009, IEEE Transactions on Nuclear Science, 56, 2351, \href{https://ui.adsabs.harvard.edu/link_gateway/2009ITNS...56.2351G/doi:10.1109/TNS.2009.2023989}{10.1109/TNS.2009.2023989}
 
  \bibitem[Godet et al. (2026b)]{Godet25} Godet, O., Atteia, J.-L., Pons, R. et al., 2026, RAA, this issue

  \bibitem[Godet et al. (2022)]{Godet22} Godet, O., Atteia, J.-L., Amoros, C. et al., 2022, Proceedings of SPIE, 12181, 121815O, \href{https://ui.adsabs.harvard.edu/link_gateway/2022SPIE12181E..5OG/doi:10.1117/12.2628932}{10.1117/12.2628932}

  \bibitem[Godet et al. (2014)]{Godet14} Godet, O., Nasser, G., Atteia, J.-L. et al., 2014, Proceedings of SPIE, 9144, 914424, arXiv:1406.7759, \href{https://ui.adsabs.harvard.edu/link_gateway/2014SPIE.9144E..24G/doi:10.1117/12.2055507}{10.1117/12.2055507}

  \bibitem[Godet et al. (2009)]{Godet09} Godet, O., Sizun, P., Barret, D. et al., 2009, Nuclear Instruments and  Methods in Physics Research A, 603, 365, \href{https://ui.adsabs.harvard.edu/link_gateway/2009NIMPA.603..365G/doi:10.1016/j.nima.2009.02.023}{10.1016/j.nima.2009.02.023}

  \bibitem[Goldwurm et (2026)]{Goldwurm25} Goldwurm, A., Bacon, P., Bellemont, N. et al., 2026, RAA, this issue

  \bibitem[Goldwurm \& Gros (2022)]{Goldwurm22} Goldwurm, A. and Gros, A., 2022, Handbook of X-ray and Gamma-ray Astrophysics. Edited by Cosimo Bambi and Andrea Santangelo, Springer Living Reference Work, 978-981-16-4544-0, arXiv:2305.10130, \href{https://ui.adsabs.harvard.edu/link_gateway/2022hxga.book...15G/doi:10.1007/978-981-16-4544-0_44-1}{10.1007/978-981-16-4544-044-1}
 
  \bibitem[G\"otz et al. (2026)]{MXT25} G\"otz, D., Crepaldi, S., Doumayrou, E. et al., 2026, RAA, this issue

\bibitem[Guillot et al. (2025)]{GCN_GRB250506A} Guillot, S., Cangemi, F., Coleiro, A. et al., 2025, GRB Coordinates Network, Circular Service, No. 40798

  \bibitem[Kirsch et al. (2005)]{Kirsch2005} Kirsch, M. G., Briel, U. G., Burrows, D. et al., 2005, Proceedings of SPIE, 5898, astro-ph/0508235, \href{https://ui.adsabs.harvard.edu/link_gateway/2005SPIE.5898...22K/doi:10.1117/12.616893}{10.1117/12.616893}


  \bibitem[Lachaud et al. (2026)]{Lachaud25} Lachaud, C., Givaudan, A., Karakac, M. et al., 2026, RAA, this issue

  \bibitem[Lacombe et al. (2013)]{L13} Lacombe, K., Nasser, G., Amoros, C. et al., 2013, Nuclear Instruments and  Methods in Physics Research A, 732, 122, \href{https://doi.org/10.1016/j.nima.2013.07.003}{10.1016/j.nima.2013.07.003}

  \bibitem[Le Stum et al. (2026)]{Lestum26} Le Stum, S., Cangemi, F., Coleiro, A. et al., 2026, \apj, 997, L25, \href{https://ui.adsabs.harvard.edu/link_gateway/2026ApJ...997L..25L/doi:10.3847/2041-8213/ae3174}{10.3847/2041-8213/ae3174} 
  
 \bibitem[Levan et al. (2025)]{Levan25} Levan, A. J., Schneider, B., Le Floc'h, E. et al., 2025, A\&A Letter, 704, L8, arXiv:2507.18783, \href{https://ui.adsabs.harvard.edu/link_gateway/2025A&A...704L...8L/doi:10.1051/0004-6361/202556581}{10.1051/0004-6361/202556581} 


 \bibitem[Lyne et al. (1993)]{Lyne_1993} Lyne, A. G., Pritchard, R. S. and Graham Smith, F., 1993, MNRAS, 265, 1003, \href{https://ui.adsabs.harvard.edu/link_gateway/1993MNRAS.265.1003L/doi:10.1093/mnras/265.4.1003}{10.1093/mnras/265.4.1003}

  \bibitem[Mate et al. (2019)]{Mate19} Mate, S., Bouchet, L., Atteia, J.-L. et al., 2019, Experimental Astronomy, 48, 171, \href{https://ui.adsabs.harvard.edu/link_gateway/2019ExA....48..171M/doi:10.1007/s10686-019-09643-x}{10.1007/s10686-019-09643-x}

  \bibitem[Piron et al. (2026)]{Piron25} Piron, F., Daigne, F., Maiolino, T. et al., 2026, RAA, this issue

  \bibitem[Qiu et al. (2026)]{VT25} Qiu, Y. L., Wang, J. M., Ho, L. C. et al., 2026, RAA, this issue

\bibitem[Sazonov et al. (2007)]{Sazonov07} Sazonov, S.,  Churazov, E., Sunyaev, R., Revnivtsev, M., 2007, MNRAS, 377, 1726, \href{https://ui.adsabs.harvard.edu/link_gateway/2007MNRAS.377.1726S/doi:10.1111/j.1365-2966.2007.11746.x}{10.1111/j.1365-2966.2007.11746.x}

  \bibitem[Schanne et al. (2026)]{Schanne25} Schanne, S., Chateau, F., Dagoneau, N. et al., 2026, RAA, this issue
  
 \bibitem[Schanne \& Godet (2024)]{Schanne24} Schanne, S. and Godet, O., 2024,  GRB Coordinates Network, Circular Service, No. 36854

 \bibitem[Schneider et al. (2026)]{Schneider26} Schneider, B. et al., 2026, submitted to A\&A, arXiv:2604.20346
 
 
    \bibitem[Shulyak et al. (2017)]{Shulyak17} Shulyak, D. Reiners, A., Engeln, A. et al., 2017, Nature Astronomy, 1, 0184, \href{https://ui.adsabs.harvard.edu/link_gateway/2017NatAs...1E.184S/doi:10.1038/s41550-017-0184}{10.1038/s41550-017-0184}

    \bibitem[Sonawane et al. (2025)]{GCNFermi} Sonawane, R., Bala S. and Meegan, C., 2025, GRB Coordinates Network, Circular Service, No. 40362

  \bibitem[Sun et al. (2026)]{GRM25} Sun, J. C., Dong, Y. W., He, J. et al., 2025, RAA, this issue

      
\bibitem[Tavani et al. (2011)]{Tavani11} Tavani, M., Marisaldi, M., Labanti, C. et al., 2011, Phys. Rev. Lett., 106, 018501, \href{https://doi.org/10.1103/PhysRevLett.106.018501}{10.1103/PhysRevLett.106.018501}

  \bibitem[Türler et al. (2010)]{Turler10} Türler, M., Chernyakova, M., Courvoisier, T. J.-L. et al., 2010, \aap, 512, 49, \href{https://ui.adsabs.harvard.edu/link_gateway/2010A&A...512A..49T/doi:10.1051/0004-6361/200913072}{10.1051/0004-6361/200913072}

    \bibitem[Wang et al. (2026)]{Wang26} Wang, J., Xie, W. J., Cangemi, F. et al., 2026, \apj, 998, 287, arXiv:2601.16558, \href{https://ui.adsabs.harvard.edu/link_gateway/2026arXiv260116558W/doi:10.48550/arXiv.2601.16558}{10.48550/arXiv.2601.16558} 


   \bibitem[Wei et al. (2016)]{Wei16} Wei, J., Cordier, B., Antier, S. et al., 2016, {\it SVOM} White Paper, arXiv:1610:06892, \href{https://ui.adsabs.harvard.edu/link_gateway/2016arXiv161006892W/doi:10.48550/arXiv.1610.06892}{10.48550/arXiv.1610.06892}

   \bibitem[Zhang et al. (2007)]{Zhang07} Zhang, B., 2007, Chinese Journal of Astronomy and Astrophysics, 7, 1, \href{https://ui.adsabs.harvard.edu/link_gateway/2007ChJAA...7....1Z/doi:10.1088/1009-9271/7/1/01}{10.1088/1009-9271/7/1/01}


     

 

\end{thebibliography}
\end{document}